# Study of spatial resolution in a single GEM simulated by Monte-Carlo method

**YANG Lan-Lan[1] (杨兰兰) , TU Yan（屠彦）, MA Shan-Le（马善乐）,ZHANG Pan-Pan（张盼盼）**

School of Electronic Science and Engineering

Southeast University, Nanjing, Jiangsu Province, P R China

**Phone**: +86(025)83792449 **Fax:** +86(025)83363222 **Email**: jujube_yang@seu.edu.cn

**Abstract:**

Spatial resolution is a significant factor in the GEM performance in view of X-rays radiography and UV, visible light imaging. Monte-Carlo method is used to investigate the spatial resolution determined by the transverse diffusion in the device. The simulation results indicate that the electrical parameters, such as the GEM voltages and the electric field at the drift and induction regions, only have minor effects on the spatial resolution. The geometrical parameters and the working gases chosen, on the other hand, are the main parameters that determine the spatial resolution. The spatial resolution is determined more on the drift and diffusion processes than on the avalanche process. Especially for the different working gases, the square root function of the ratio of the electron diffusion coefficient and the mobility has a significant effect on the spatial resolution.

**Key words:** Gas electron multiplier；Spatial resolution；Monte-Carlo Method

**PACS** 29.40.Cs

## 基于蒙特卡罗方法的 GEM 位置分辨率研究

杨兰兰*，屠彦，马善乐，张盼盼

电子科学与工程学院，东南大学，南京，中国

**摘要：**

基于气体电子倍增器（GEM）在 X 射线、UV 光和可见光成像领域的应用，位置分辨率成为 GEM 探测器的一个重要性能指标。本文采用蒙特卡罗方法研究位置分辨率的特性。研究结果表明，GEM 膜间电压、漂移区和扩散区电场等一些电参数对位置分辨率的影响很小，而几何形状参数和工作气体对位置分辨率影响较大，可以看出漂移扩散过程远较倍增过程对位置分辨率的影响大。特别是针对不同工作气体而言，$\sqrt{D/K}$（电子的扩散系数和迁移系数比值的平方根）对位置分辨率具有决定性的影响。

**关键词：**

气体电子倍增器；位置分辨率；蒙特卡罗方法

**PACS** 29.40.Cs

[1] Sponsored by the National Natural Science Foundation of China under Contract (No.50907009, 61271053, 60871015), the Natural Science Foundation of Jiangsu Province under Contract No. BK2012737, the Foundation for Excellent Youth Scholars of Southeast University.

# 1. Introduction

The gas electron multiplier (GEM) has already been widely used in high energy physics experiments, and is also being developed in view of photon detection and imaging applications[1-3]. The GEM is a robust device for the proportional amplification of electrons released in a gas by X-ray radiation, charged particles or light. It consists of drift region, GEM foil, induction region and PCB readout electrodes. GEM foil is a thin polymer foil, with two-side copper-clad applied by suitable voltage and perforated by a high density of apertures. In view of possible applications as large area visible light imagers, spatial resolution is now becoming a significant parameter in GEM characteristics. Spatial resolution is defined as the minimum distance between two objects which can be distinguished. It might be related to the GEM voltage, the electric filed and the distance of the drift region and induction region, GEM hole geometry, working gas and readout method. The effects of readout method or geometries of readout electrode have been well discussed in the previous articles[4,5]. In this paper, a Monte-Carlo method is used to investigate the drift and diffusion motions of the particles during the drift, amplification and collection processes, and then the spatial resolution is determined as a function of the above parameters.

# 2. GEM Simulation Cell Structure

The basic component of a GEM is a metal-coated thin insulating foil, chemically pierced with a high density of holes, typically 50-100μm in diameter at 100-200μm pitch. The polymer used for manufacturing is Kapton, 50μm thick, with 5μm copper cladding on both sides[6]. Fig.1 shows the schematic representation of the GEM simulation cell structure. The GEM foil hole geometry is a typical double-conical shape, where the standard parameter D(the outer diameter) is 70μm , d(the inner diameter) is 50μm and p(pitch) is 140μm. The lengths of drift and induction regions: $L_d$ and $L_i$ are reduced to 200 μm in order to keep one crucial parameter avoid other disturbances, although the actual size in a real structure is several millimeters. The reduction of the drift and induction regions is acceptable because the electric fields out of 200μm are almost uniform. The voltage difference between the drift electrode and the upper copper electrodes is 40V, so the drift electric field strength $E_d$ can be roughly estimated as 2kV/cm. The induction electric field strength $E_i$ is determined by the 60V voltage difference between the readout electrode and the lower copper electrode, so that $E_i$ is approximately 3kV/cm. Several hundred volts $V_{GEM}$ are applied across the two copper electrodes on both sides of the GEM Kapton foil. The basic periodic element(rectangle layout) of the GEM foil geometry used in the simulation is shown in Fig.1(b).

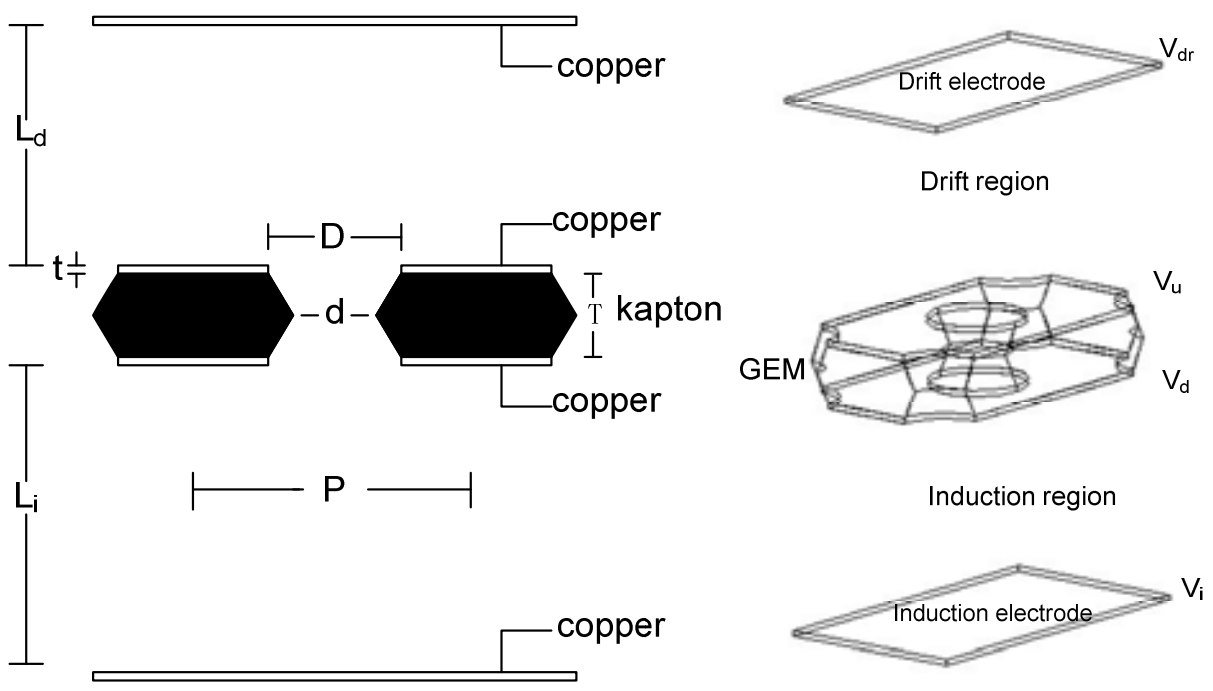

(a) (b)

Fig. 1 Schematic representation of the GEM simulation structure (a) GEM geometries and materials of different parts (b) GEM simulating structure model and applied voltages

## 3. Description of the Monte-Carlo Method

The Garfield program is used in the simulation, which is developed for gaseous detectors by Rob Veenhof at CERN[7]. The calculation procedure of the GEM simulation is shown in Fig.2(a) and the detailed Monte-Carlo method to simulate the particle drift, diffusion and avalanche is shown in Fig.2(b). In order to simulate the GEM characteristics, the finite element method is applied to generate grids based on different boundaries such as electrode, dielectric, periodic boundary and nearly arbitrary boundary shapes. Then the electrostatic field and magnetic field are calculated according to the Maxwell equations. In addition, the single gas or gas mixture properties, including particle transport properties are calculated and embedded in the program. After that, the Monte-Carlo integration method is used to calculate the particle drift, diffusion and avalanche from the starting point until the particle ends in the electrode or other boundaries. As shown in Fig.2(b), step length and diffusion coefficients are generated. And then the mean gas multiplication over a drift path is estimated as the exponential of the integral of the Townsend coefficient over the path:

$$M = e^{\int \alpha dz} \qquad (1)$$

The integration is performed using the Newton-Raphson technique over each step of the drift line, repeatedly bisecting a step[7]. By this method, the electron avalanche is simulated along the drift path. Finally, all the particles have been simulated, and the spatial resolution can be obtained from the charges position collected at the readout electrode. The spatial resolution is defined as the standard deviation of the spread of the electrons distribution at the readout electrode.

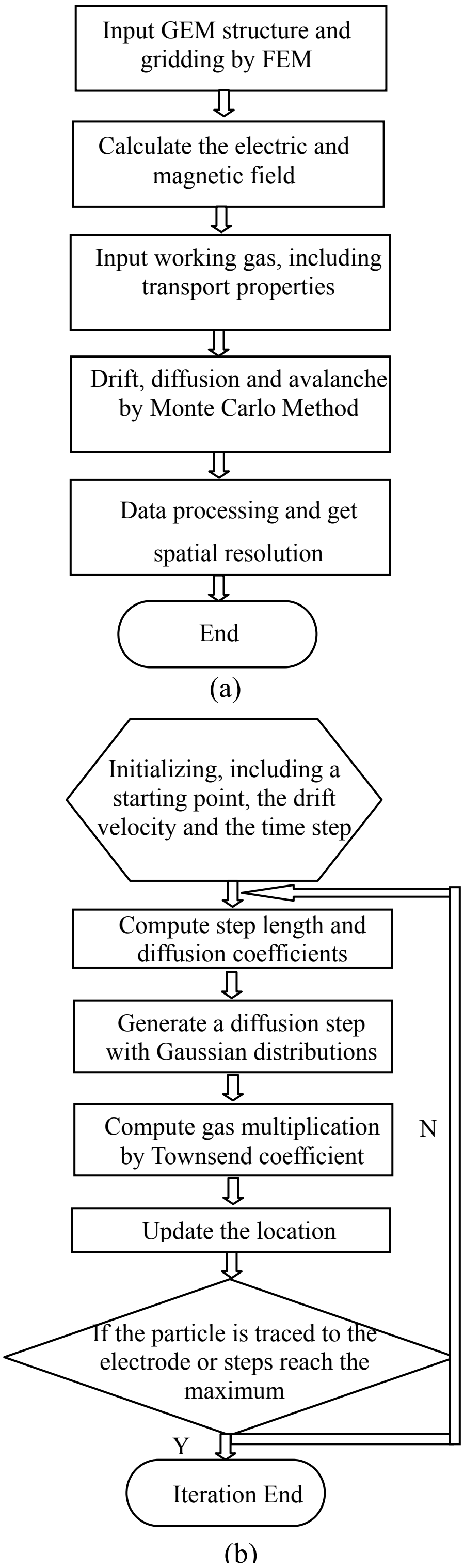

Fig. 2 Calculation procedure of GEM simulation (a) GEM simulation procedure (b) calculation procedure of Monte-Carlo integration method

Fig.3 (a) illustrates the single electron multiplication process. Fig.3(b) shows electrons distribution at the readout electrode in GEM after the multiplication of 1000 initial electrons with an angle perpendicular to the surface. Fig.3(c) shows the Gaussian fit of the electron distribution at the induction electrode. Fig.3 (b)(c) are obtained by simulating 1000 initial electrons, emitted at x=370μm spot. And we get xc=370.837μm where xc is the mean or expectation of the distribution. That is to say, the imaging spot corresponds to the emitting spot when the emitting position is along the center line of the hole. Fig.3(d) shows the expectation value at different emitting spots. It seems that the maximum deviation of the expectation of the distribution from the emitting position is around 18μm, which is one eighth of the hole pitch(140μm). From Fig.3, we can see that the normal distribution fits well for the electron distribution at the induction electrode. In a GEM, while a particle drifts, on average it will follow the drift velocity vector. Additionally, it will be scattered transversely. The transverse diffusion will make a particle follow a trajectory that differs from the mean. The spread by diffusion is calculated by propagating a probability distribution along the drift path. Transverse diffusion is obtained due to convergence of drift lines neighboring to the central drift line, which is considered Gaussian at every stage. The transverse diffusion is the decisive factor to the spatial resolution. Finally, the spatial resolution, that is to say, the total diffusion at the induction electrode, also obeys the Gaussian distribution, as proven by the simulation and experimental results[8]. The electron distribution, i.e., charge position, is collected by the readout(induction) electrode to generate the detection signal. Excluding the influences of the readout electrode's geometry and signal reading method, the GEM's working gas, geometric shape and electrical parameter significantly influence the spatial resolution of the GEM. Although there are several definitions of resolution according to different criteria, in fact all of them are equivalent[9]. As the electron distribution obeys the normal distribution, in this paper we use the standard deviation σ as the spatial resolution, as this ultimately limits the resolving power of imaging detectors. The dependence of spatial resolution as a function of working gas, GEM geometric shape and electrical parameter is important to GEM device and is calculated in this paper. In this paper, when investigating the effect of one parameter, we keep the other parameters unchanged if there are no special declarations.

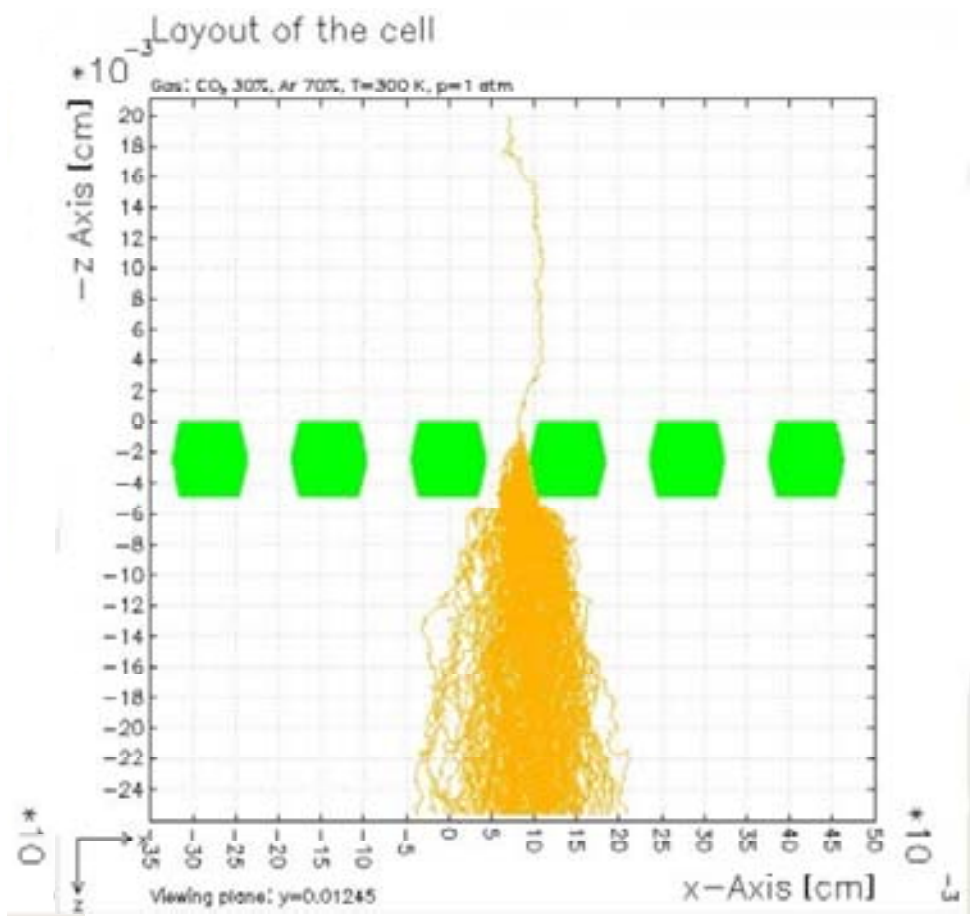

(a)

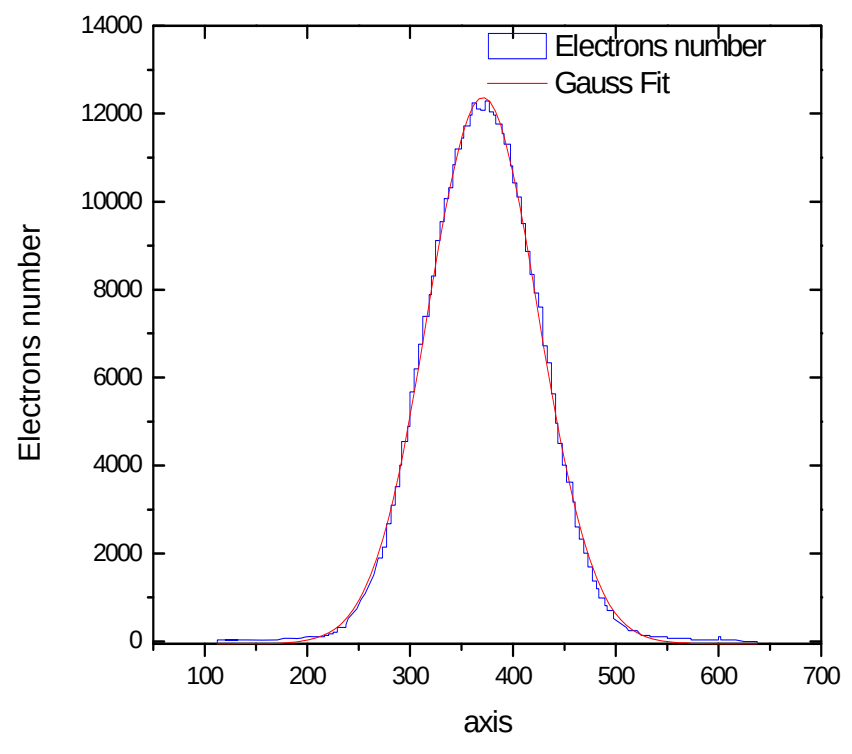

(b)

| Model | Gauss | | |
|---|---|---|---|
| Equation | y=y0 + (A/(w*sqrt(PI/2)))*exp(-2*((x-xc)/w)^2) | | |
| Reduced Chi-Sqr | 52088.58404 | | |
| Adj. R-Square | 0.99742 | | |
| | | Value | Standard Error |
| Electrons number | y0 | -49.82801 | 31.08535 |
| | xc | 370.83745 | 0.20496 |
| | w | 107.34287 | 0.53815 |
| | A | 1.67077E6 | 10136.88053 |
| | sigma | 53.67144 | |
| | FWHM | 126.38658 | |
| | Height | 12418.93549 | |

(c)

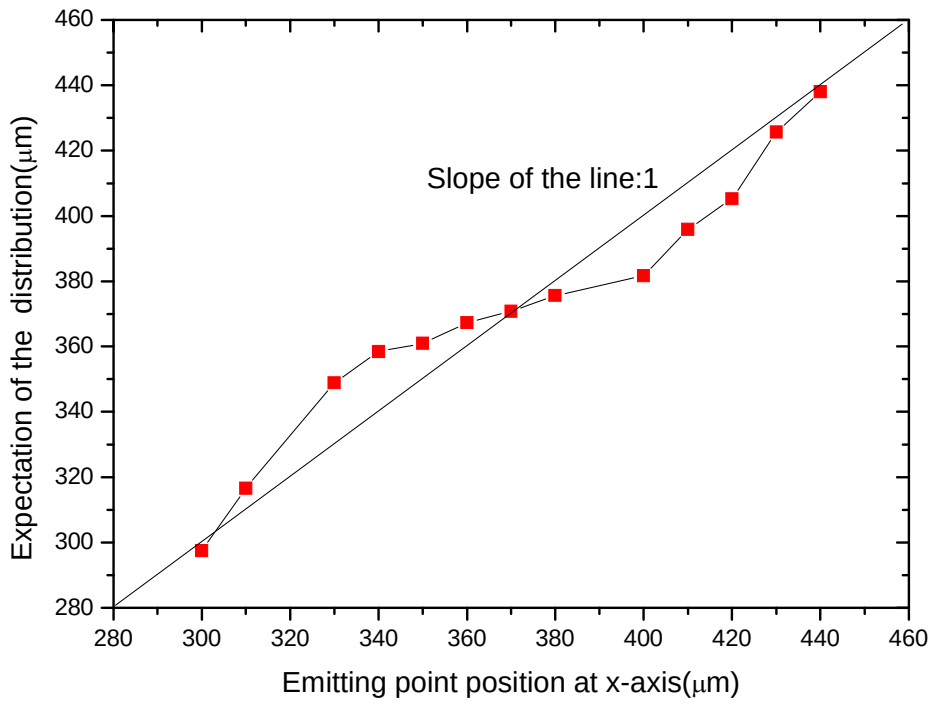

(d)

Fig.3 Description of the electron multiplication process and the electron distribution at the induction electrode (a) single electron multiplication process (b) electron distribution collected by the induction electrode (c) Gaussian fit of the electron distribution (d) Expectation value of the distribution with different emitting spot

## 4. Simulation Results and Discussions

### 4.1 Effects of GEM Electrical Parameters

As we know, GEM voltages have great effects on the GEM gain, while the drift and the induction electric field also influence the GEM gain. While concerning the spatial resolution, these electrical parameters only have minor effects on the spatial resolution. Fig.4 shows the spatial resolution at different induction electric field strengths and GEM voltages, where x-axis and y-axis respectively describe the resolution data at the x-axis direction and y-axis direction. As the structure is identical at both x and y directions, the difference is merely showing the statistical error. Here the effects of the drift electric field are omitted due to the similarity to the induction electric field. From Fig.4, we can see that the spatial resolution varies around 5μm when the induction electric field strength varies from 1KV/cm to 7KV/cm, or when the GEM voltages vary from 300V to 500V. That is to say, the electrical parameters such as GEM voltages and the electric field at the drift and induction regions, are not decisive factors to the spatial resolution.

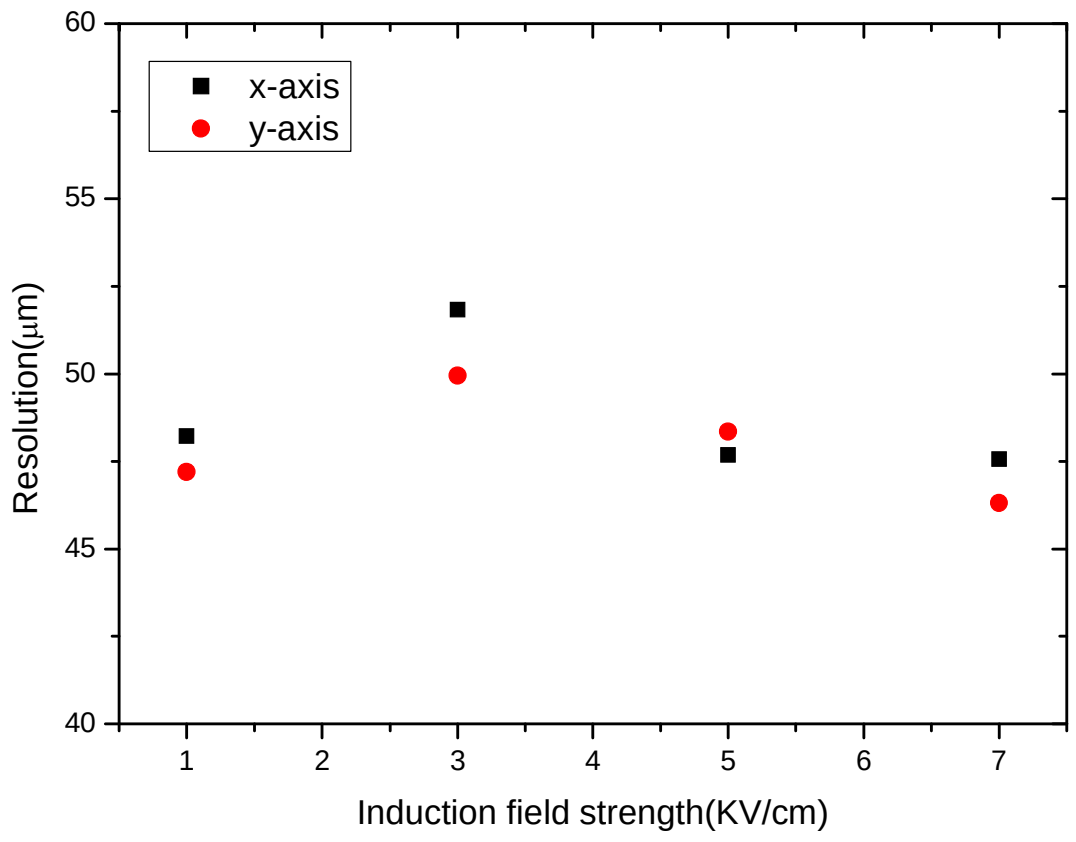

(a)

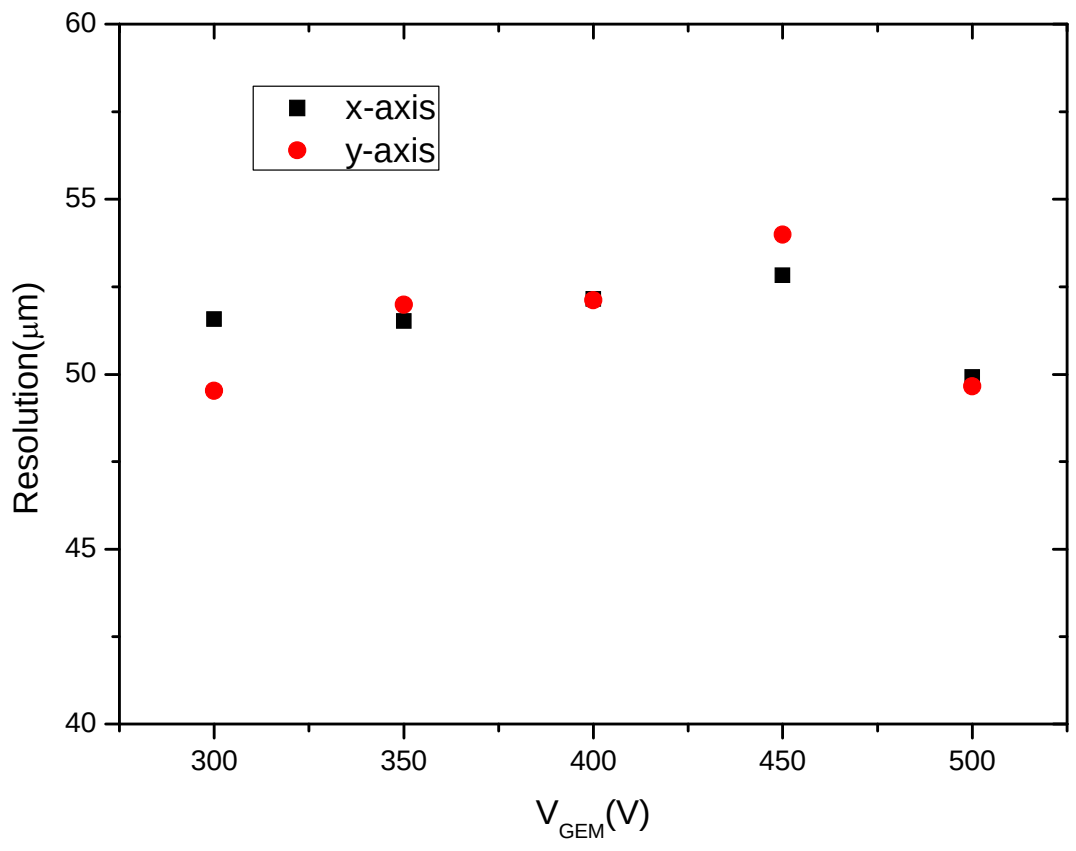

(b)

Fig.4 Spatial resolution at different induction electric field strengths and GEM voltages (a) Induction electric

field strengths (b) GEM voltages

### 4.2 Effects of GEM Geometrical Parameters

In this section, the effects of the drift region length, the induction region length, the GEM hole diameter and GEM hole geometry to the spatial resolution are investigated. Fig.5 shows the spatial resolution as a function of the drift region length $L_d$ and the induction region length $L_i$. Fig.5 (a)(b) tell us that the influences of the $L_d$ and $L_i$ to the spatial resolution are similar to each other, where both of them obey a square root function $\sigma=2.76L_d^{0.5}$ and $\sigma=2.76L_i^{0.5}$.

If we assume GEM as a cylindrical device at the z-direction electric field, the electrons can transverse average distance r according to diffusion theory in gas discharge[10], where

$$r = \sqrt{\frac{4Dz}{u_e}} \qquad (2)$$

In this formula, D, z, $u_e$ represent the diffusion coefficient, the drift distance and the electron velocity respectively.

Therefore, we can conclude that the spatial resolution is similar to the transverse average distance, and has a square root function with the drift distance, both in the drift region and the induction region. According to Formula(2), we calculate the proportional coefficient is 2.87 and 3.37 which corresponds to the above fitting coefficient 2.76 and 2.76. In addition, the simulated tendency is consistent with the experimental results in Ref[5].

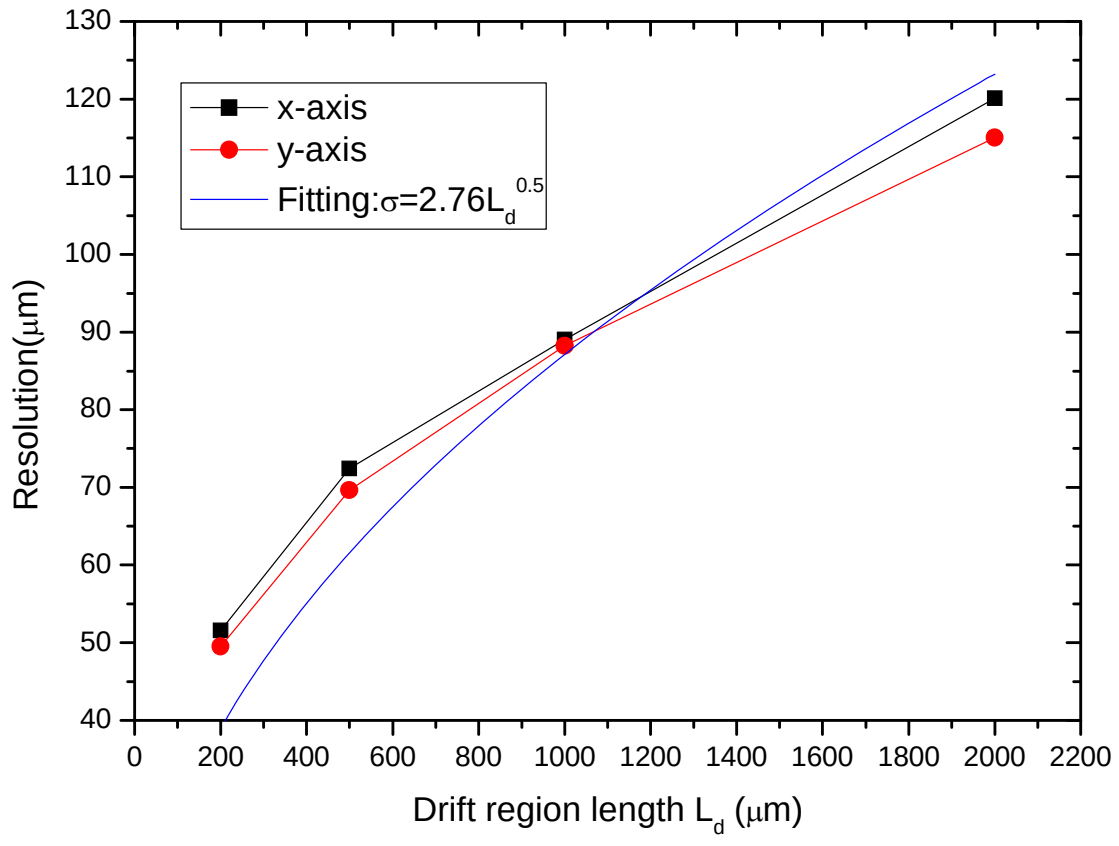

(a)

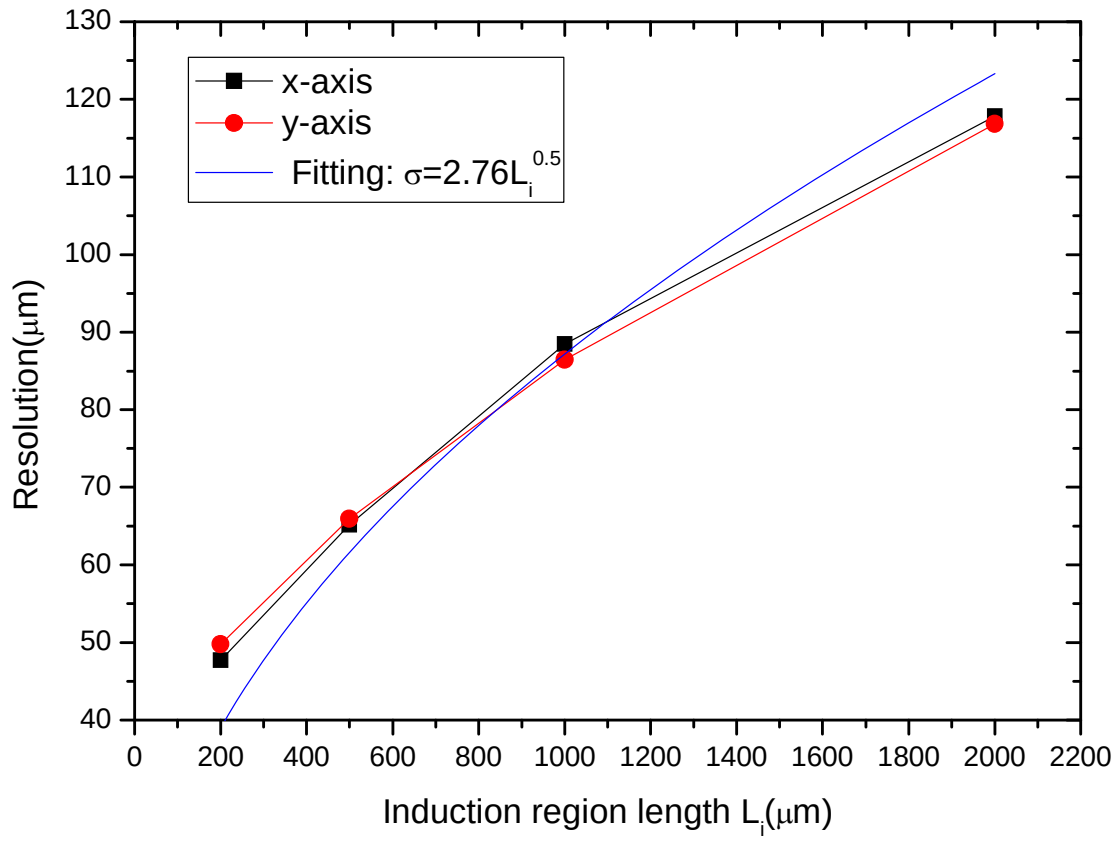

(b)

Fig.5 Spatial resolution varies with the drift region and induction region lengths (a) drift region length (b) induction region length

Fig.6 shows the spatial resolution varies with the GEM hole diameter, using a cylindrical GEM hole shape. It can be seen that the resolution increases linearly with the GEM hole diameter. The fitting formula is: $\sigma=29.83+0.28D_h$. Fig.7 shows that the spatial resolution varies with different types of hole geometrical structure: double conical, cylindrical, single conical, reverted-conical. $D_1$, $D_2$, $D_3$ respectively represent the diameters at different positions of four types of structures. In Fig.7(b), $D_1$-$D_2$-$D_3$ at horizontal ordinate can describe the hole geometry. From Fig.7(b), we can see that $D_3$,i.e. the diameter at the GEM bottom layer, is the significant factor to the spatial resolution.

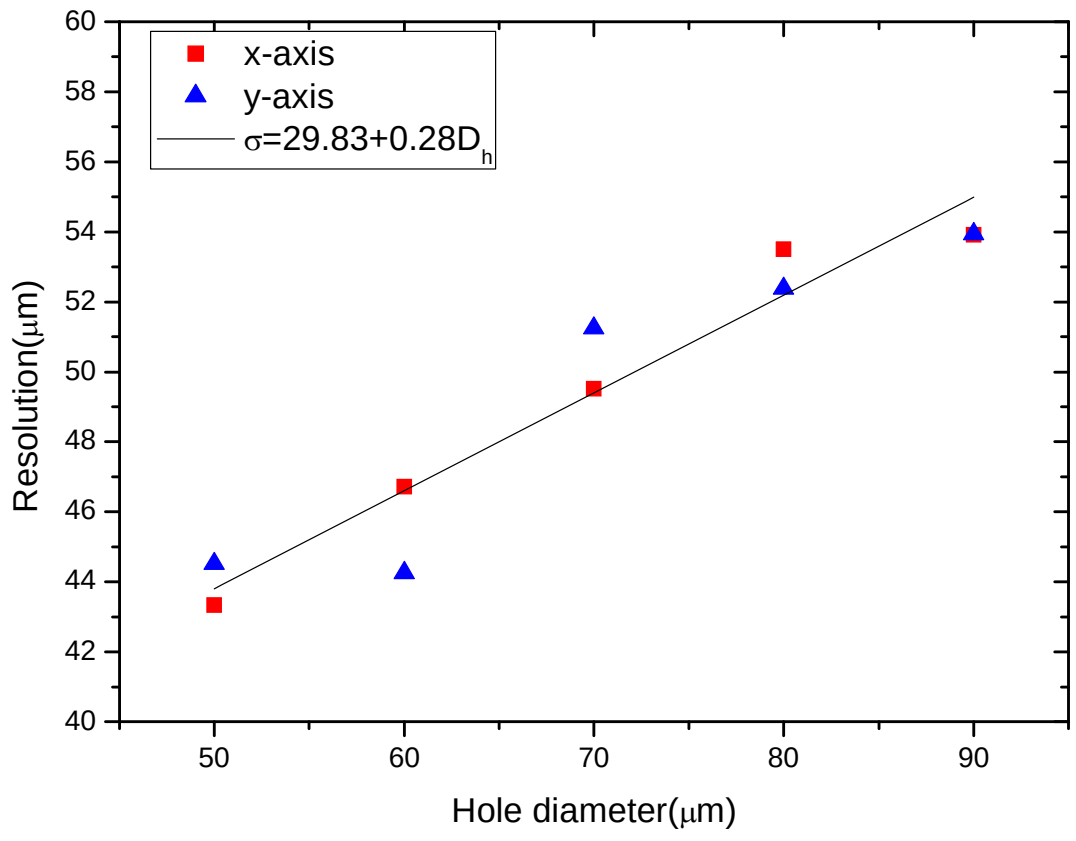

Fig.6 Spatial resolution vary with GEM hole diameter

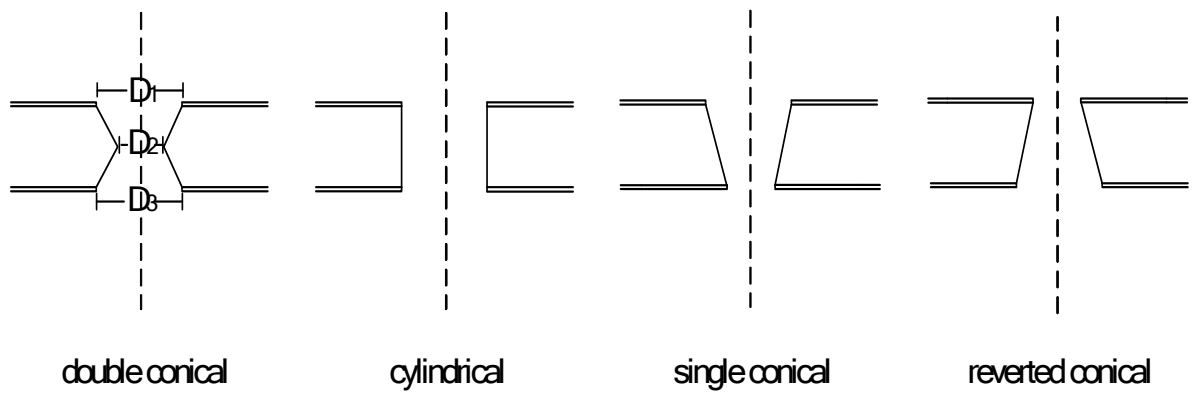

(a)

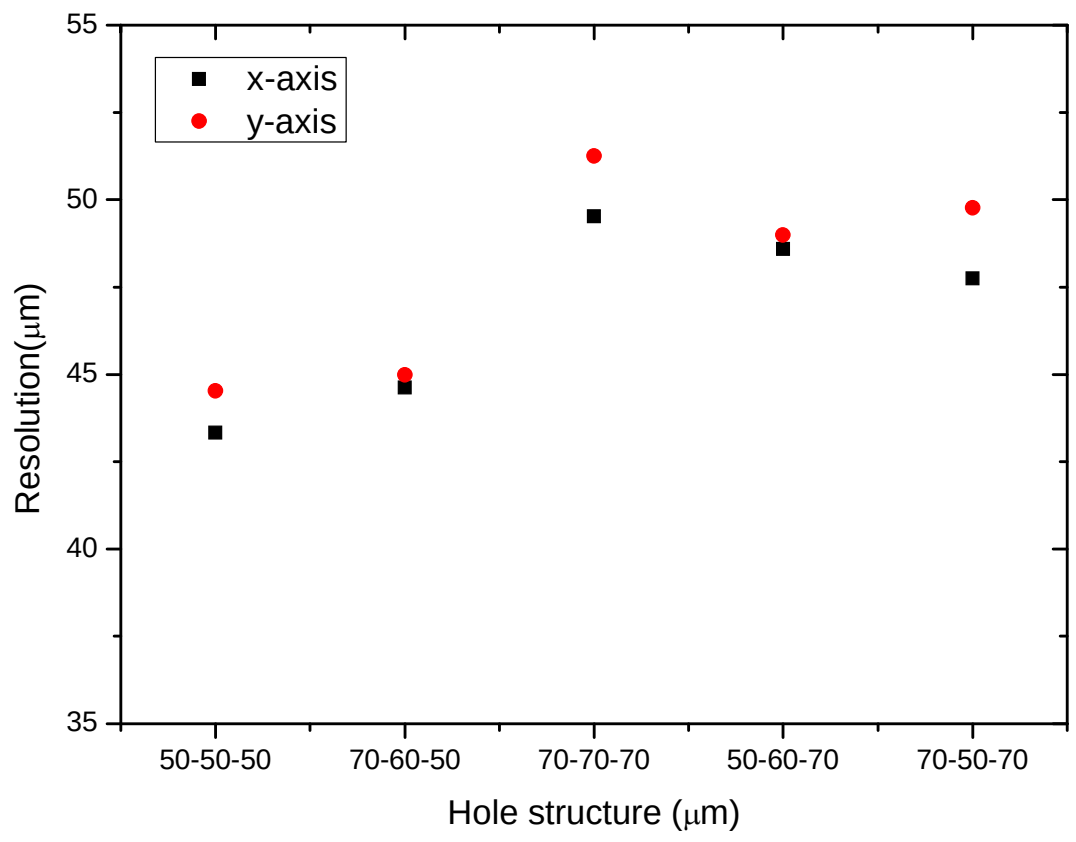

(b)

Fig.7 Spatial resolution as a function of the GEM hole geometrical structure (a) Description of different types of hole geometrical structure (b) Relation between spatial resolution and hole structure

### 4.3 Effects of Different GEM Working Gases

Fig.8 shows the spatial resolution when using different working gases. The spatial resolution increases with the increase of $CO_2$ contents. As $CO_2$ is a type of quenching gases, it has larger cross sections compare to Ar. While $CO_2$ contents keep at 30%, the spatial resolutions vary a little with the change of the background gas: Ar, Ne, and Xe. That tells us that $CO_2$ plays a very import role to the spatial resolution for the above reason. When the background gas Ar is kept at 70%, the spatial resolution of $CH_4$ is larger than $CO_2$. $CH_4$ is a fast gas with large diffusion while $CO_2$ is a slow gas with relatively small diffusion. Therefore, the transverse diffusion will spread wider in $CH_4$ than in $CO_2$. According to Formula(2), the average transverse distance can be rewritten:

$$r = \sqrt{\frac{4Dz}{u_e}} = \sqrt{\frac{4Dz}{K \cdot E}} \propto \sqrt{\frac{D}{K}} \quad (3)$$

where K, E represent respectively the mobility of electrons and the electric field strength.

Fig.9 shows $\sqrt{D/K}$ of different gas types at different E/N values(N is the concentration of gas neutral particles). E/N is usually expressed using the Townsend unit, written as "Td", which equals to $10^{-17}$ V.cm$^2$. The value of the mobility K and the diffusion coefficient D are obtained from BOLSIG[11,12]. It can be clearly seen that the shape of spatial resolution is similar to the $\sqrt{D/K}$ shape at around several tens of Td when the other conditions are kept the same. Considering the electric field strengths variation range at the drift region, avalanche region and the induction region, the reduced electric field is rudely varied from 1~250Td. Fig.10 shows the normalization of the $\sqrt{D/K}$ value and the spatial resolution. It can be seen that spatial resolution in most of the cases corresponds to the $\sqrt{D/K}$ value between 10Td and 20Td, which is slightly above the drift and the induction electric field strength. Therefore, the $\sqrt{D/K}$ value is the most decisive factor to the spatial resolution of the different working gases.

In conclusion, the spatial resolution depends significantly on the drift and the diffusion process, which is related to the drift and induction region lengths, the working gases. The avalanche process, which is related to the GEM voltages, has only a small effect on the spatial resolution. The GEM hole geometry shape, particularly the diameter of the GEM bottom layer is also the significant factor to the spatial resolution.

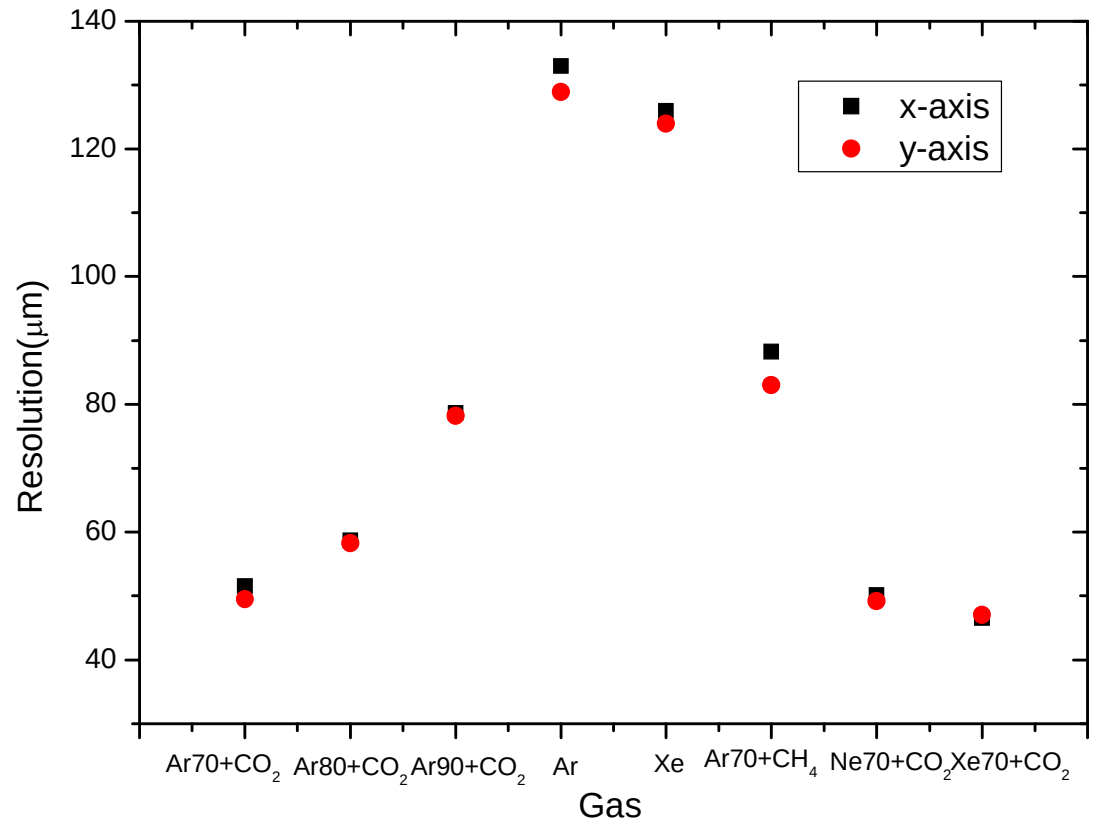

Fig.8 Spatial resolutions at different working gases

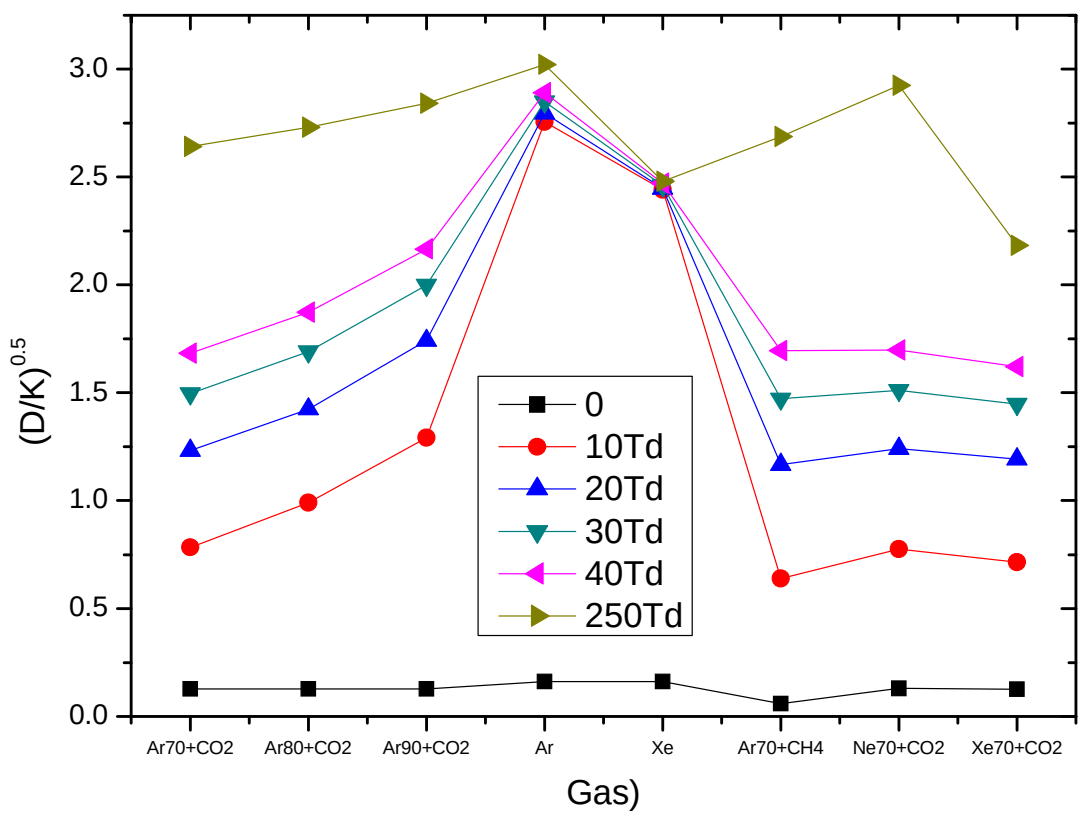

Fig.9 $\sqrt{D/K}$ of different gas types at different E/N values.

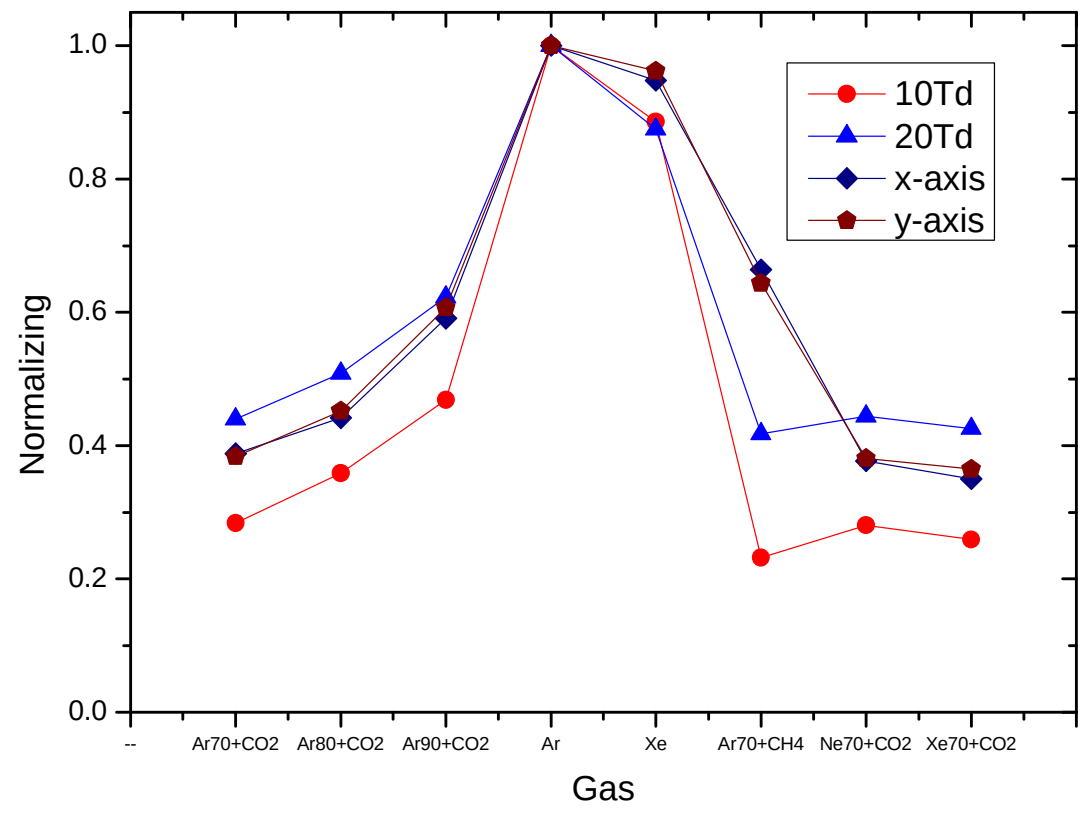

Fig.10 Normalization of the $\sqrt{D/K}$ value and the spatial resolution

## 5. Conclusions

In view of its use in soft X-rays radiology and large area visible light imagers, the spatial resolution is now becoming a significant factor in GEM characteristics. Garfield program with Monte-Carlo method is used in the simulation to investigate the spatial resolution. Excluding the influences of the readout electrode's geometry and the signal reading method, GEM's working gas, geometric shape and electrical parameter significantly influence the spatial resolution of GEM. The electrical parameters such as GEM voltages and the electric field at the drift and induction regions are not decisive factors to the spatial resolution. However, the geometrical parameters and the working gases influence the spatial resolution significantly. We can conclude that the spatial resolution depends mostly on the drift and diffusion processes, not the avalanche process in the GEM. The spatial resolution has a square root dependence on the drift distance, both in the drift region and the induction region. And it depends linearly on the diameter of GEM holes. In particular, the diameter of at the GEM bottom layer is the significant factor to the spatial resolution.

For different working gases, the spatial resolution is proportional to the square root of the diffusion coefficient divided by the mobility. The results obtained, are expected to be of considerable help to optimize the GEM design with respect to the spatial resolution in the different applications. Nevertheless more theoretical and experimental investigations should be carried out for a full study of the spatial resolution in GEM, such as the influence of the track angle, the reading method and the beam intensity distribution.